\documentclass[10pt, letter,twocolumn]{IEEEtran}
\pdfoutput=1

\usepackage[dvips]{color}
\usepackage{epsf}
\usepackage{times}
\usepackage{epsfig}
\usepackage{graphicx}
\usepackage{epstopdf}
\usepackage{algorithm}
\usepackage{algorithmic}
\usepackage{amsmath}
\usepackage{amssymb}
\usepackage{amsxtra}
\usepackage{multirow}
\usepackage{mathtools}
\usepackage{cite}   
\usepackage{color}
\usepackage{bbm}
\usepackage{subfigure}
\usepackage{nomencl}
\usepackage{hyperref}
\title{{\fontsize{23}{30}\selectfont A Latency-Aware Task Offloading in Mobile Edge Computing Network for Distributed  Elevated LiDAR}} 

\author{
\IEEEauthorblockN{\large Michael C. Lucic$^1$, Hakim Ghazzai$^1$, Ahmad Alsharoa$^2$, and Yehia Massoud$^1$}\\
\IEEEauthorblockA{\small $^1$School of Systems and Enterprises -- Stevens Institute of Technology, Hoboken, NJ, USA}\\ %Email: %\{mlucic, hghazzai, ymassoud\}@stevens.edu}\\
\IEEEauthorblockA{\small $^2$Missouri University of Science and Technology, Rolla, MO, USA\\ %Email: aalsharoa@mst.edu\\
}
{\thanks {\hrule
\vspace{0.1cm} \indent This paper is accepted for publication in the IEEE International Symposium on Circuits and Systems (ISCAS '20), Seville, Spain, Oct. 2020. 

\textcopyright~2020 IEEE.  Personal use of this material is permitted.  Permission from IEEE must be obtained for all other uses, in any current or future media, including reprinting/republishing this material for advertising or promotional purposes, creating new collective works, for resale or redistribution to servers or lists, or reuse of any copyrighted component of this work in other works.
}}
}

\makenomenclature
\begin{document}
\maketitle
\thispagestyle{empty}
\begin{abstract}
\boldmath
Recently, elevated LiDAR (ELiD) has been proposed as an alternative to local LiDAR sensors in autonomous vehicles (AV) because of the ability to reduce costs and computational requirements of AVs, reduce the number of overlapping sensors mapping an area, and to allow for a multiplicity of LiDAR sensing applications with the same shared LiDAR map data. Since ELiDs have been removed from the vehicle, their data must be processed externally in the cloud or on the edge, necessitating an optimized backhaul system that allocates data efficiently to compute servers. In this paper, we address this need for an optimized backhaul system by formulating a mixed-integer programming problem that minimizes the average latency of the uplink and downlink hop-by-hop transmission plus computation time for each ELiD while considering different bandwidth allocation schemes. We show that our model is capable of allocating resources for differing topologies, and we perform a sensitivity analysis that demonstrates the robustness of our problem formulation under different circumstances. \vspace{-0.1cm} 
\end{abstract}
\begin{IEEEkeywords}
Elevated LiDAR, intelligent transportation system, mobile edge computing, optimization. \vspace{-0.4cm}
\end{IEEEkeywords}
\section{Introduction}
\label{Sec1}

Light detection and ranging (LiDAR) technology has taken root as a core sensing technology in various Intelligent Transportation Systems (ITS) applications, including, but not limited to, autonomous vehicles (AV) \cite{Luettel2012} and topological urban 3D mapping \cite{Sun2013}. In AVs, LiDAR point cloud maps are combined with camera-based images and Radio Detection and Ranging (RADAR) detection data for the purpose of Simultaneous Locating and Mapping (SLAM) of AVs \cite{Masmoudi2019a,Cadena2016}, which is used to aid the AV control systems in navigating through dynamic surroundings in real-time~\cite{Masmoudi2019b}. 

While LiDAR provides AVs with a rich source of data for aiding in its SLAM solution and navigation protocols, there are a few major issues with placing LiDAR sensors in AVs. First, LiDAR sensors are very expensive. For example, a high rotation speed Velodyne LiDAR commonly used in AVs may retail for around \$8,000 (a lower cost figure is expected if mass-produced) \cite{Gunnam2018}. In the event that multiple LiDARs are equipped on an AV (such as an autonomous tractor-trailer truck), the cost of the LiDAR sensors alone could easily end up being as much as the rest of the vehicle itself, providing an enormous barrier to entry and distribution of AV technology to the average user. Second, LiDAR sensors require large amounts of on-board power-hungry components, such as graphics processing units (GPU), to process, store, and fuse the LiDAR data with other on-board sensor data in real-time \cite{Venugopal2013}. This can lead to drastic fuel efficiency losses in the operation of AVs. According to a Bloomberg article in \cite{Coppolla2017}, AVs have 10\% worse fuel efficiency than human-operated vehicles due to the huge power demands from on-board sensors and processing units, further increasing costs for consumers and burdens on the environment. Third, a LiDAR on-board an AV has a much worse perspective of its surroundings than a LiDAR in an elevated position, due to the reduced field-of-view (FoV) angle versus an aerial perspective. In the literature, the majority of LiDAR mapping applications actually consider an aerial perspective (i.e. LiDARs mounted on aircraft \cite{Sun2013}), that can increase the sensing range, and therefore requiring less sensors. 

To counter the pitfalls of on-board LiDAR in AVs, a novel Elevated LiDAR (ELiD) system has been proposed~\cite{Jayaweera2019}. The ELiD system counters the high cost, inefficiency, and low FoV perspective of AV-based LiDARs and GPUs. The idea is to move the LiDARs to elevated positions (either lamp-posts or sides of buildings), and the GPUs to the cloud or on the edge. The ELiDs are connected to roadside units (RSUs)~\cite{Lucic2019a,Lucic2019b} that may dynamically shift coverage~\cite{Lucic2019c} to transfer ELiD's data to/from the AVs. 
\begin{figure}[t!]
\centering
\vspace{0.2cm}
% \begin{tabular}{cc}
\includegraphics[width=7cm]{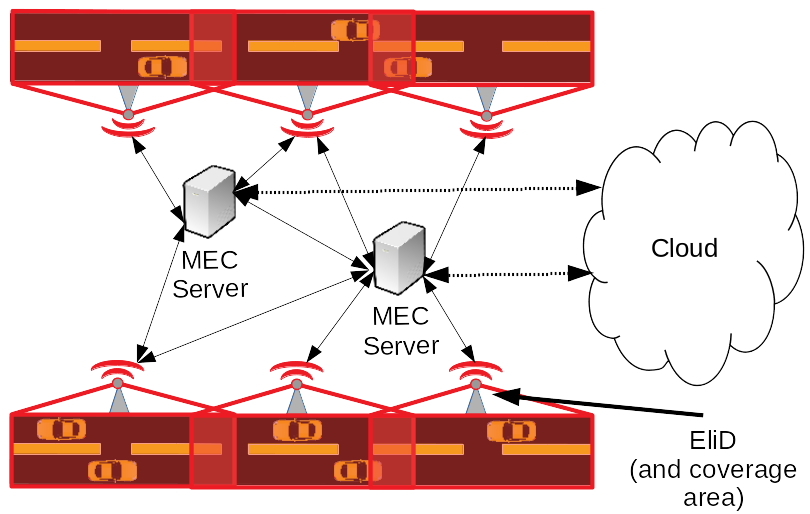}\vspace{-0.4cm}
% \end{tabular}
\caption{An example network topology, where ELiD devices are scanning the roads, and sending data to MEC devices, and data exceeding the MEC device limits are sent to the cloud.}\vspace{-0.5cm}
\label{systemModel}
\end{figure}
In addition to the primary use for AV SLAM, the elevated perspective of the LiDARs provides the capability for the ELiD system to perform mapping tasks for urban municipal authorities, such as infrastructure monitoring. For example, many cities around the world (e.g. Mexico City, Mexico and Beijing, China, among others) are sinking from excess groundwater extraction \cite{Parker2016}. The ELiD system can gain fine-grained data on which areas are sinking the fastest, and which areas are flood-prone because of changes in topography.

One of the main challenges in ELiD system implementation is is the need for a backhaul management system \cite{Jayaweera2019} to route ELiD data from the sensors to servers (edge or cloud) via existing network connections. In this paper, we propose an efficient solution to this challenge. The goal is to minimize total network latency, as LiDAR sensing is a mission-critical task in AVs for safety reasons \cite{Lin2018}. Therefore, we develop and solve an mixed integer optimization problem that minimizes the latency of the backhaul network that connecting the ELiD sensors with mobile edge computing (MEC) devices, while taking into account the servers' processing speed and memory resources. We propose to use GPU's or custom field-programmable gate arrays (FPGA) \cite{Lyu2018} with the cloud for processing the ELiD-generated map. The model aims to consider RAM constraints of the edge and cloud servers. The latency is based on two parts: 1) the hop-by-hop latency of the uplink and downlink transmissions, and the latency associated with the time needed to process the data in the MEC devices/cloud. Our mixed-integer programming problem solution will optimally (i.e. minimal latency) decide on how bandwidth is shared across the MEC backhaul network.
\vspace{-0.15cm}

\section{System Model}
\label{Sec2}
We consider a backhaul network that consists of a set of nodes $\mathcal{N}$, including the set of ELiD's $\Lambda$, and networked computers $\mathcal{S}$, such that $\mathcal{N} = \Lambda \cup \mathcal{S}$. Fig. \ref{systemModel} illustrates an example of the network topology. The networked computers in our problem are assumed to be either routers, MEC devices, or the cloud. Routers act as nexus points to route communications and make connections with other points in the network without processing the data. MEC devices are smaller, less powerful computers that are located physically close to the ELiD's. The cloud is much more powerful than the MEC decices. Data may be sent to the cloud if the data processing requires more resources than whats available at the edge. We assume all nodes are connected by fiber-optic cables. Individual ELiDs are denoted by the index $\lambda \in \Lambda$, individual servers are denoted by $s \in \mathcal{S}$, and any node can be indexed by $i, j\in \mathcal{N}$.  

We define the binary parameter $e_{ij}$ to indicate whether or not nodes $i$ and $j$ connected by a fiber link, where $e_{ij} = 1$
if the nodes $i$ and $j$ are connected with a fiber-optic link and 0 otherwise. An existing fiber-optic link between nodes $i$ and $j$ has an average total bandwidth of $R_{ij}$ bytes/s. The fiber-optic links are bidirectional, meaning that uplink and downlink signals are transmitted over different visual light spectra.

\begin{figure}[t!]
\centering
\begin{tabular}{cc}
\includegraphics[width=5cm]{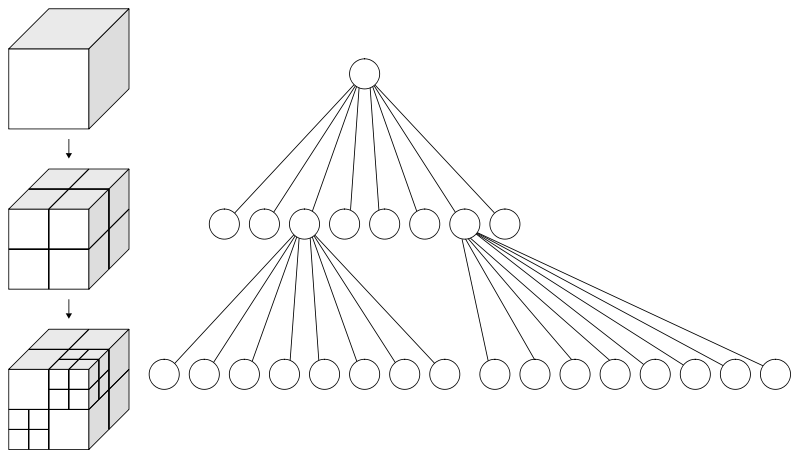} &\includegraphics[width=3cm]{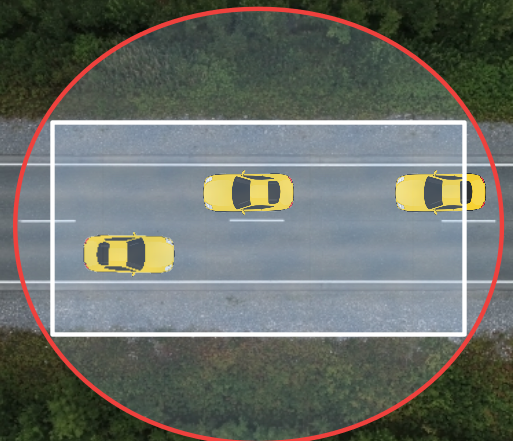}
\end{tabular}\vspace{-0.2cm}
\caption{(Left) Illustrating the octree data structure used for LiDAR mapping. (Right) The ELiD coverage area is marked with the red circle. The white box shows the scan boundary caused by buildings running along the road.}
\label{octree}\vspace{-0.3cm}
\end{figure}

The scanning areas of the ELiDs depend on the horizontal/vertical FoV angles. In the problem, we assume that we are optimizing the ELiD backhaul for an urbanized area. Due to the buildings that are placed close to the roadway, we assume the ELiD coverage sector would be roughly in the shape of a rectangular prism due to the obstruction from the buildings (see Fig. \ref{octree}). The data structure of the ELiDs' data is an octree. In essence, the scan sector is divided, and if an object is detected in that area, it is further subdivided into a smaller set, recursively ($d$ times). The granularity of the scan is of a specified precision in $2^{1 - d}$ m. If no objects are in a subdivided area, it is left empty, and the depth of the subdivisions is not increased in that sector, in order to efficiently store and process data \cite{Kumar2012}. 

Assuming a cubic scan sector with a volume of 1 m$^3$, the octree contains at most $D_{oct}$ bytes/m$^3$ of data per scan that is given as:
\begin{equation}
D_{oct} = 8^{(d-2)} + 12,
\label{scanSize}
\end{equation}
where the first term in \eqref{scanSize} represents the number of bits of data to map to the depth of the recursion $d$ multiplied by the number of bytes per bit contained in each layer. Please note the size of the data depends on the resolution of the scan. The second term in \eqref{scanSize} equals 12 bytes, which corresponds to the 3-tuple of 32-bit (4 byte) floating point numbers that represents the reference coordinate for the scan area.

We assume that LiDAR sensors perform a full scan each rotation. Thus, the produced data rate can be given as follows:
\begin{equation}
D_{\lambda} = D_{oct} \times f_{scan} \times V_{scan} \times \Gamma,
\end{equation}
where $D_{\lambda}$ is the data production rate from ELiD $\lambda$ in bytes/s, $f_{scan}$ is the number of times ELiD $\lambda$ rotates per scan, and $\Gamma$ is the compression factor~\cite{Anand2019}.

We assume a processed map $D_{s}$ returning from a server $s$ contains less data then the raw scan data $D_{\lambda}$. We model this size difference as $\beta = \frac{D_{s}}{D_{\lambda}} \leq 1$.

The coverage areas of ELiDs are non-homogeneous, as a result of different traffic flow rates, road topographies, and other characteristics that make some areas more important to cover than others. We therefore define a priority score $\rho_{\lambda}$ for each ELiD $\lambda$, where a lower value corresponds to a more important priority.

Servers in the network have characteristics to be accounted for. A computer $s$ in the network has $M_{s}$ bytes of random access memory (RAM) that limits the number of jobs running on the server. In addition to RAM, the GPU affects processing throughput. Since ELiD's produce image-like data, GPU's are effective for processing LiDAR data. GPU benchmarks aid in estimation of the data in bytes processed per second in the server~$\omega_s$.

% \vspace{-0.3cm}

\section{Problem Formulation}
\label{Sec3}
In this section, we formulate a mixed-integer non-linear programming (MINLP) problem aiming to minimize the total ELiDs' backhaul latency.

\subsection{Objective Function}
\label{Sec3A}
Our objective is to route network traffic to/from the ELiDs to servers and process the data in a timely manner, while ensuring effective data routing. We define the round-trip plus processing latency of a data transmission $\mathcal{L}_{\lambda}$ as follows:

\begin{subequations}
\begin{align}
\hspace{-0.5cm}\mathcal{L}_{\lambda} &= \left[\sum_{i \in \mathcal{N}} \sum_{j \in \mathcal{N}} \Upsilon_{ij}^{\lambda} + \Delta_{ij}^{\lambda}\right] + \sum_{s \in \mathcal{S}} P_{s}^{\lambda}, \forall \lambda \in \Lambda, \\
\text{where }&\Upsilon_{ij}^{\lambda} = u_{ij}^{\lambda} \frac{D_{\lambda}}{\alpha_{ij}R_{ij}}, \forall \lambda \in \Lambda, \forall i, j \in \mathcal{N}, \\
&\Delta_{ij}^{\lambda} = \delta_{ij}^{\lambda} \frac{D_{s}}{\gamma_{ij}R_{ij}}, \forall \lambda \in \Lambda, \forall i, j \in \mathcal{N}, \\
&P_{s}^{\lambda} = c_{s}^{\lambda} y_{s} \frac{D_{\lambda}}{\omega{s}}, \forall \lambda \in \Lambda, \forall s \in \mathcal{S}.
\end{align}
\label{LambdaRef}
\end{subequations}

In \eqref{LambdaRef}, $\Upsilon_{ij}^{\lambda}$ and $\Delta_{ij}^{\lambda}$ represent hop-by-hop transmission latencies for the uplink and downlink, respectively, and $P_{s}^{\lambda}$ is the expected processing time of the data based on server hardware benchmarks. Additionally, $u_{ij}^{\lambda}$ and $\delta_{ij}^{\lambda}$ are binary decision variables that indicate the uplink or downlink message from/to $\lambda$ that passes from $i$ to $j$. The variables $\alpha_{ij}$ and $\gamma_{ij}$ represent the percentage of the fiber-optic bandwidth allocated to uplink and downlink signals, respectively, between two nodes $i$ and $j$, where $\alpha_{ij} + \gamma_{ij} \leq 1$ and $\alpha_{ij}, \gamma_{ij} \in [0, 1]$. Regarding the server processing latency, $c_{s}^{\lambda}$ denotes whether or not the data from $\lambda$ being processed in server $s$. Finally $y_{s}$ represents the number of jobs being processed simultaneously in server $s$. We assume that if $n$ jobs are running in a server, then a job therefore the can take $n$ times longer to complete, because LiDAR point cloud map processing is embarrassingly parallel due to the tensor-like structure of the data. Therefore, our objective function is given as:
\begin{equation}
L_{func} = \sum_{\lambda \in \Lambda}\rho_{\lambda}\mathcal{L}_{\lambda}.
\label{objFun}
\end{equation}

\subsection{Problem Constraints}
\label{Sec3B}
The total latency objective function given in \eqref{objFun} is constrained by the following:

\subsubsection{Node Flow Constraints}
\label{Sec3B1}
The message (uplink or downlink) originating at LiDAR $\lambda$ arriving from a node $i$ to server or router $s$ must be either departed from server $s$ to node $j$ or be processed in the server. Therefore the following constraints need to be respected:
\begin{subequations}
\begin{align}
\sum_{i \in \mathcal{N}} u_{is}^{\lambda} &= \sum_{j \in \mathcal{N}} u_{is}^{\lambda} + c_{s}^{\lambda}, \forall \lambda \in \Lambda, \forall s \in \mathcal{S}, \\
\sum_{i \in \mathcal{N}} \delta_{si}^{\lambda} &= \sum_{j \in \mathcal{N}} \delta_{js}^{\lambda} + c_{s}^{\lambda}, \forall \lambda \in \Lambda, \forall s \in \mathcal{S}.
\end{align}
\label{firstConstr}
\end{subequations}  

\subsubsection{Uplink/Downlink Requirements}
\label{Sec3B2}
All LiDARs must send their data to be processed, and on the other hand receive a processed map back. In addtion, LiDARs do not act as routers:
\begin{equation}
\sum_{j \in \mathcal{N}} u_{\lambda j}^{\lambda} = 1, \sum_{j \in \mathcal{N}} \delta_{j\lambda}^{\lambda} = 1, \forall \lambda \in \Lambda.
\label{uldlreq}
\end{equation}  

\subsubsection{Network Topology Limitations}
\label{Sec3B3}
We assume that the uplink and downlink segments along the network occur only where a fiber-optic link exists:
\begin{equation}
% \begin{align}
u_{ij}^{\lambda}, \delta_{ij}^{\lambda} \leq e_{ij}, \forall i, j \in \mathcal{N}, \forall \lambda \in \Lambda.
 % &\leq e_{ij}, \forall i, j \in \mathcal{N}, \forall \lambda \in \Lambda.
% \end{align}
\label{topreq}
\end{equation}  

\subsubsection{Prevention of Double-Counting}
\label{Sec3B4}
We assume that the uplink and downlink segments occur only in one direction on a link between nodes $i$ and $j$:
\begin{equation}
% \begin{align}
(u_{ij}^{\lambda} + u_{ji}^{\lambda}), (\delta_{ij}^{\lambda} + \delta_{ji}^{\lambda}) \leq 1, \forall i, j \in \mathcal{N}, \forall \lambda \in \Lambda.
 % \leq 1, \forall i, j \in \mathcal{N}, \forall \lambda \in \Lambda.
% \end{align}
\label{dcreq}
\end{equation}   

\subsubsection{Server RAM Constraints}
\label{Sec3B5}
We assume that server $s$ processing capacity is constrained by its RAM capacity. We assume no read/write operations on the storage disk, due to the high latency of I/O operations:
\begin{equation}
\sum_{\lambda \in \Lambda} D_{\lambda}c_{s}^{\lambda} \leq M_{s}, \forall s \in \mathcal{S}.
\label{ramreq}
\end{equation}     

\subsubsection{Defining Number of Active Jobs on a Server}
\label{Sec3B6}
The number of jobs running on a server $s$ is based on the number of LiDARs that send data to server $s$:
\begin{equation}
y_{s} = \sum_{\lambda \in \Lambda} c_{s}^{\lambda}, \forall s \in \mathcal{S}.
\label{jobreq}
\end{equation}    

\subsubsection{Job Processing Requirements}
\label{Sec3B7}
All LiDARs must have their data processed in a server $s$:
\begin{equation}
\sum_{s \in \mathcal{S}} c_{s}^{\lambda} = 1, \forall \lambda \in \Lambda.
\label{lastConstr}
\end{equation}

\subsection{Optimization Problem and Bandwidth Allocation}
\label{Sec3C}
Based on the objective function formulated in Section \ref{Sec3A} and the constraints formulated in Section \ref{Sec3B}, we can formulate our MINLP optimization problem as follows:
\begin{align*}
(\mathcal{P}_X)\min\limits_{\mathcal{L}_{\lambda}}&\sum_{\lambda \in \Lambda}\rho_{\lambda}\mathcal{L}_{\lambda}\\
\text{s.t. }&\text{Constraints \eqref{LambdaRef}, \eqref{firstConstr} - \eqref{lastConstr}, and ($Z$),}
\end{align*}
Since $\alpha_{ij}$ and $\gamma_{ij}$ are in the denominator of the objective function, $(\mathcal{P}_0)$ is intractable. Therefore, in order to find a tractable solution, we must redefine $\alpha_{ij}$ and $\gamma_{ij}$ to utilize pre-existing numerical methods to solve this problem. In the subsequent sections, we discuss three methods for redefining these variables in order to make the problem tractable.

\subsubsection{Fixed Bandwidth Allocations}
\label{Sec3D}
If we predefine channels of a certain fraction of the total fiber-optic bandwidth size $\epsilon$, then we have the following:
\begin{equation}
\alpha_{ij} = \gamma_{ij} = \epsilon, \forall i, j, \in \mathcal{N}.
\label{fixedfrac}
\end{equation}

\subsubsection{Decoupled Variable Allocations}
\label{Sec3E}
If the downlink message sizes are some percentage $\beta$ of the uplink message sizes, then we may allocate $\sigma = \frac{1}{1 + \beta}$ percent of the channel to the uplink, and $1 - \sigma$ percent to the downlink. This would result in the following:
\begin{equation}
% \begin{align}
\frac{1}{\alpha_{ij}} = \frac{\sum_{\lambda \in \Lambda}u_{ij}^\lambda}{\sigma}, \frac{1}{\gamma_{ij}} = \frac{\sum_{\lambda \in \Lambda}\delta_{ij}^\lambda}{1 - \sigma}, \forall i, j \in \mathcal{N}, \\
% , \forall i, j \in \mathcal{N},
% \end{align}
\label{decoupled}
\end{equation}
where we have equalities because minimizing latency is the dual problem of maximizing the utilization of the network capacity.
\begin{figure}[t!]
\centering
\begin{tabular}{cc}
\includegraphics[width=4cm]{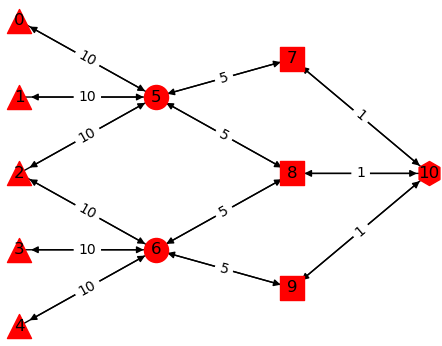} & \includegraphics[width=4cm]{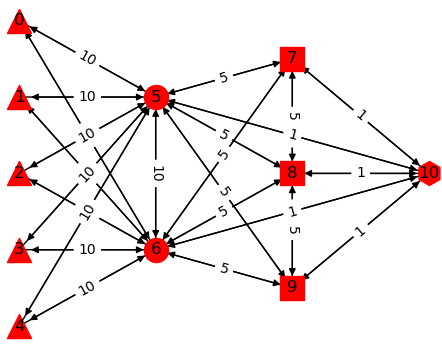}
\end{tabular}
\caption{The two network topologies used in testing our problem, with the same number of nodes $\mathcal{N}$. The triangles represent ELiD units, the circles represent routers, the squares represent MEC devices, and the hexagon represents the cloud. The labels on the edges are the link bandwidths in GB/s.}
\label{iniTopology}\vspace{-0.4cm}
\end{figure}

\subsubsection{Combined Variable Allocations}
Taking a combined approach from Sections \ref{Sec3D} and \ref{Sec3E}, we assume that uplink or downlink transmissions can be allocated some variable percentage of the channel size $\epsilon_{ij} = \alpha_{ij} = \gamma_{ij}$. This would lead to the following:
\begin{equation}
\frac{1}{\epsilon_{ij}} = \sum_{\lambda \in \Lambda}u_{ij}^{\lambda} + \delta_{ij}^{\lambda}, \forall i, j \in \mathcal{N},
\label{together}
\end{equation}
where again we have an equality because minimizing latency is the dual problem of maximizing the utilization of the network capacity.

\subsubsection{New mixed integer quadratically constrained problem (MIQCP) formulations}
As a result of the assumptions made in the previous few subsections, our objective function becomes quadratic in all three cases. Therefore, we can now formulated our general optimization problem $(\mathcal{P}_X)$ as the following MIQCP: 
\begin{align*}
(\mathcal{P}_X)\min\limits_{\mathcal{L}_{\lambda}}&\sum_{\lambda \in \Lambda}\rho_{\lambda}\mathcal{L}_{\lambda}\\
\text{s.t. }&\text{Constraints \eqref{LambdaRef}, \eqref{firstConstr} - \eqref{lastConstr}, and ($Z$),}
\end{align*}
where we have the three problems $(\mathcal{P}_1)$ if $Z = \eqref{fixedfrac}$, $(\mathcal{P}_2)$ if $Z = \eqref{decoupled}$, and $(\mathcal{P}_3)$ if $Z = \eqref{together}$ based on which definition for $\alpha_{ij}$ and $\gamma_{ij}$ are used. Since in all three formulations we have MIQCPs, we may use an off-the-shelf solver such as Gurobi \cite{Gurobi}, which takes advantage of state-of-the-art integer programming numerical methods such as branch-and-bound, cutting planes, and model presolve to determine the optimal solution of our problem formulations.

\begin{figure}[t!]
\centering
\begin{tabular}{cc}
\includegraphics[width=4cm]{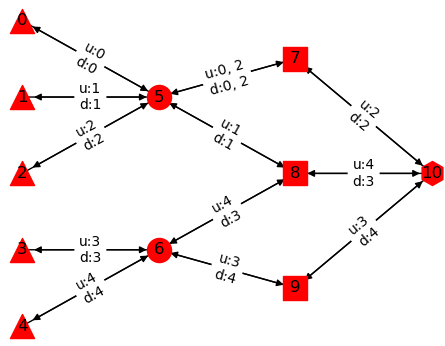} & \includegraphics[width=4cm]{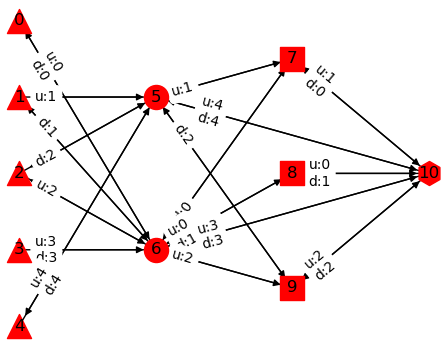}
\end{tabular}
\caption{The solutions for the network topologies presented in Figure \ref{iniTopology}, given the initial model parameters in Table \ref{paramTable}. The solution on the left has an average processing time of 384ms, where the solution on the right has an average processing time of 284ms.}
\label{solTopology}\vspace{-0.4cm}
\end{figure}

\section{Results and Discussion}
\label{Sec4}

In this section, we provide example solutions (Figure \ref{solTopology}) for two network topologies (sparse and dense, see Figure \ref{iniTopology}) for $(\mathcal{P}_3)$, perform a brief sensitivity analysis given parameter and constraint variations on the sparse network topology, and interpret our results.
% Sensitivity Analysis of D_l vs Latency given variation in CPU processing abilities of edge servers
In the initial model runs, we illustrate how the network traffic is allocated across both the sparse and dense topologies using the $(\mathcal{P}_3)$ formulation, given the model parameters in Table \ref{paramTable}. The parameters were based on the technical specifications and benchmarks of the hardware assumed to be incorporated in the system, where we consider NVIDIA Jetson GPU system-on-chip (SoC) as the MEC devices, and we assume that the cloud instance for our network has NVIDIA Tesla V100 GPUs on-board. We see a shorter average latency in the dense network solution, as traffic has more distribution paths across the network.
\begin{table}[H]
\centering
\caption{\label{paramTable} Initial Model Run Parameters}
\vspace{-0.2cm}
\addtolength{\tabcolsep}{-4pt}\begin{tabular}{|c|c|c||c|c|c|}
\hline
\textbf{Parameter} & \textbf{Value} & \textbf{Unit} & \textbf{Parameter} & \textbf{Value} & \textbf{Unit} \\
\hline
$D_{\lambda}$ & 100 & MB/s & $\beta$ & 0.8 & n/a\\
\hline
$R_{ij}$ & 1, 5, 10 & GB/s & $\rho_{\lambda}$ & $\lambda$ & n/a\\
\hline
$\omega_{s}$ & 0.25, 55 & GB/s & $M_{s}$ & 1, 256 & GB\\
\hline
\end{tabular}\vspace{-0.3cm}
\end{table}

\begin{figure}[t!]
\centering
% \hspace{-2cm}
% \begin{tabular}{cc}
\includegraphics[width=8.75cm]{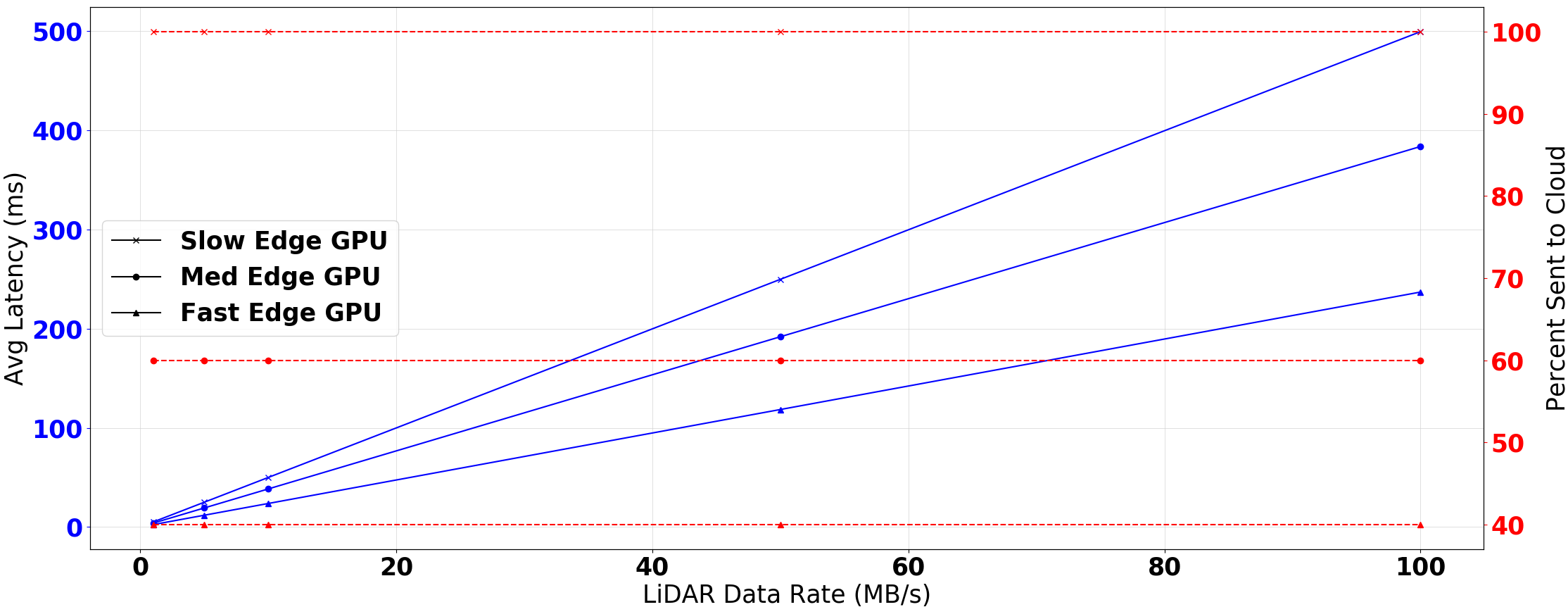}\vspace{0.1cm} \includegraphics[width=8.75cm]{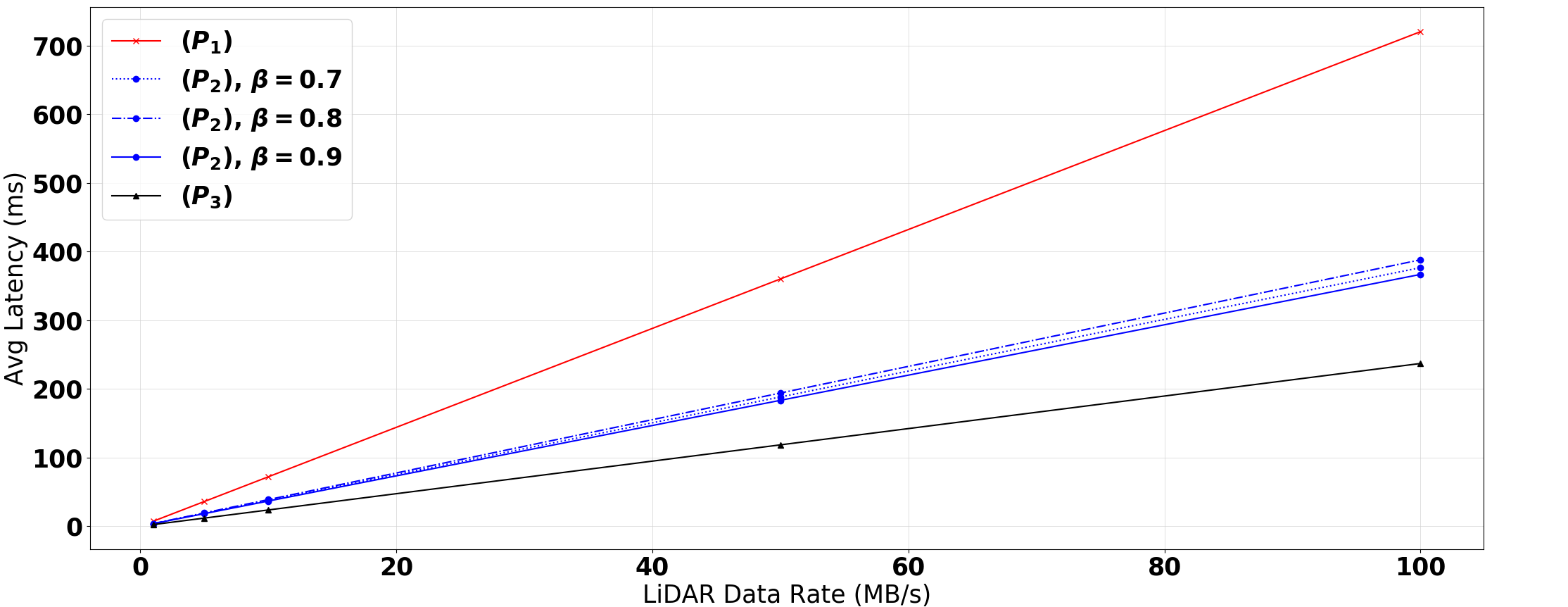}
% \end{tabular}
\vspace{-0.2cm}
\caption{Plots of our sensitivity analysis on different aspects of the model. The top plot looks at how the ELiD sensing resolution affects the average latency for the system and the percentage of the messages sent to the cloud given different edge server processing speeds. The bottom plot looks at how different bandwidth-sharing constraints affect the LiDAR sensing resolution vs. latency relationship.}
\label{SensAnalysis}\vspace{-0.4cm}
\end{figure}

In our sensitivity analysis on the sparse network setting, we investigate the relationship between $D_{\lambda}$ and the average latency to process data from a ELiD $\lambda$ given slow (50MB/s), medium (250MB/s), and fast (1GB/s) processing speeds $\omega_s$ for the edge servers, as well as how much traffic must be sent all the way to the cloud. We see a linear increase in average latency with an increase in $D_{\lambda}$. In addition, we see that as the bases with more powerful edge servers, less data was sent to the cloud. Please see Figure \ref{SensAnalysis}, top pane, for an illustration of this. For example, we see that if the ELiD units produce 50 MB/s of compressed data, The slow MEC device processing speed leads to around 250ms of average total latency, the medium speed MEC (the approximate benchmark speed of the NVIDIA Jetson SoC for image processing), we have an average latency of just under 200ms, and if the MEC devices had GPU's 4x faster then the Jetson SoC the average latency is just above 100ms. We also see in this cases that the MEC devices are faster, less data is sent to the cloud, showing the trade-off between the transmission latency and the processing latency. Additionally, the last value in the medium MEC device processing speed curve in Figure \ref{SensAnalysis} corresponds to the result in our initial model run for the sparse network (Figure \ref{solTopology}, left side).

Afterward, we evaluate the performance of $(\mathcal{P}_1)$, $(\mathcal{P}_2)$, and $(\mathcal{P}_3)$ given variations in $D_{\lambda}$. From the bottom pane in Figure \ref{SensAnalysis}, we again see the positive linear relationship between average latency and $D_{\lambda}$, where $(\mathcal{P}_1)$ performs the worst, $(\mathcal{P}_2)$, and $(\mathcal{P}_3)$ performs the best, because equation \eqref{together} in $(\mathcal{P}_3)$ provides more efficient allocation of bandwidth to the uplink/downlink channels. In another example, in the case of $D_{\lambda} = 50$ MB/s, we see that when we fix the percentage of the channel allocated to a message to be $\epsilon = 10$\%, we have an average latency of approximately 350ms, versus just under 200ms for all of the decoupled allocations, and just over 100ms for the combined allocation.

\section{Conclusion}
\label{Sec5}
In this paper, we developed a series of MIQCPs capable of routing ELiD data to MEC and cloud servers in a way to minimize latency for a fixed-topology fiber-optic based backhaul network. Moving forward, we plan to develop a more computationally efficient algorithm that may scale to much larger networks, so this system may be utilized in a real-life ELiD setting. In addition, we aim to consider splitting data across different paths during transmission to further maximize the utilization of the network.

\bibliographystyle{ieeetr}

\end{document}